\documentstyle[preprint,aps]{revtex}
\tighten

\begin{document}

\bibliographystyle{unsrt}

\draft

\title{{\normalsize{\rm
NORDITA 94/69 \hfill gr-qc/9412053 \\
{\mbox{ }} }}\\
Dark matter and non-Newtonian gravity from General Relativity
coupled to a fluid of strings}

\author{Harald H. Soleng\footnote{
E-Mail address: Soleng@surya11.cern.ch}\footnote{Present address:
Theory Division, CERN, CH-1211 Geneva 23, Switzerland}}

\address{NORDITA, Blegdamsvej 17,
         DK-2100 Copenhagen {\O}, Denmark}

\date{September 27, 1993; revised August 18, 1994}

\maketitle

\begin{abstract}
An exact solution of Einstein's field equations for
a point mass surrounded by
a static, spherically symmetric
fluid of strings
is presented.
The solution is singular at the origin. Near
the string cloud limit
there is a $1/r$ correction to Newton's force law.
It is noted that at
large distances and small accelerations, this
law coincides with the phenomenological
force law invented by Milgrom in order to explain
the
flat rotation curves of galaxies without introducing dark matter.
When interpreted
in the context
of a cosmological
model with a string fluid,
the
new solution
naturally explains why the
critical acceleration of Milgrom is of the same order of magnitude
as the Hubble parameter.
\end{abstract}

\pacs{PACS numbers: $\;$ 04.20.-q $\;$ 98.80.-k\\
{\mbox{ }}\\
To appear in the Journal of General Relativity and Gravitation}

\renewcommand{\theequation}{\arabic{section}.\arabic{equation}}
\renewcommand{\thefootnote}{\arabic{footnote}}
\addtocounter{footnote}{-\value{footnote}}

%%%%%%%%%%%%%%%%%%%%%%%%%%%%%%%%%%%%%%%%%%%%%%%%%%%%%%%%%%%%%%%%%%%

\section{Introduction}
\addtocounter{equation}{-\value{equation}}

According to the string paradigm
the fundamental building blocks of nature are extended objects
of dimension one. It is therefore important to understand the
gravitational fields produced by a collection of strings rather
than point particles. In this vein the concepts of dust
clouds and perfect fluids
have been extended to  {\em string clouds\/}
\cite{Letelier} and perfect
{\em string fluids\/} \cite{Letelier2}.
The string-dominated universe model discussed by Vilenkin \cite{Vilenkin84}
corresponds to a universe filled with a string cloud.
Another approach has been to find exact solutions
of gravitational field equations in the presence of
a finite number of strings.
Examples are the solution for a straight cosmic string
\cite{Gott2,Hiscock}, a solution for
two moving straight strings \cite{Gott}, and a solution for
$N$ straight
strings moving on a circle \cite{Kabat}.

Letelier \cite{Letelier2} found exact integral expressions for
the general solution to Einstein's
field equations coupled to a fluid of strings with spherical symmetry.
Here we shall consider
a static, spherically symmetric
string fluid where the
transverse
pressure
is proportional to the energy density.
In this case, we are able to give explicit expressions for the
exact solution. It is found
that a point particle
surrounded by a string fluid with a small transverse
pressure
produces a gravitational field corresponding to
a short distance $1/r^2$ and a long distance $1/r$ force law.
As in the vacuum case, the solution is
singular at the origin, but this does not affect the validity of the
solution as a model of the gravitational field at large $r$.

As an application of the new solution, a string fluid cosmology is presented.
In this model the universe is filled with a net of
strings. It is assumeed that
these strings are pulled out of the net
by galaxies and become
nearly radially threaded through the galaxies.
It is argued that this hypothetical scenario
can explain the flat rotation curves of galaxies.

\section{Fluid of strings}
\addtocounter{equation}{-\value{equation}}

For the sake of completeness
a short review of
the results of Letelier
\cite{Letelier,Letelier2}
is given
in this section.
Just as we have the four-velocity, $u^{\mu}$, associated with
the world line of a particle, we have the
bivector, $\Sigma^{\mu\nu}$, associated
with the world sheet of a string.
Let the string world sheet be coordinated by $\lambda^{A}\in \{\lambda^{0},
\lambda^{1}\}$, where $\lambda^{0}$ and $\lambda^{1}$ are timelike and
spacelike coordinates,
respectively. Then the world sheet forming bivector is given as
\begin{equation}
\Sigma^{\mu\nu}=\epsilon^{AB}\frac{\partial x^{\mu}}{\partial\lambda^{A}}
\frac{\partial x^{\nu}}{\partial\lambda^{B}}
\label{bivector}
\end{equation}
where $\epsilon^{AB}$ is the antisymmetric,
two-dimensional Levi-Civita symbol.

The energy-momentum tensor for a dust cloud is given as
$T_{c}^{\mu\nu}=\rho u^{\mu}u^{\nu}$ where
$u^{\mu}$
is the
normalized four-velocity
($u^{\mu}u_{\mu}=-1$)\ of a dust particle.
Similarly, for a string cloud
one defines
\cite{Letelier}
\begin{equation}
T_{c}^{\mu\nu}=\rho\sqrt{-\gamma}
\Sigma^{\mu\lambda}\Sigma_{\lambda}^{\;\;\nu}/(-\gamma)
\end{equation}
where
$\Sigma^{\mu\nu}$ is the world sheet
bivector of each string and where
\begin{equation}
\gamma\equiv\frac{1}{2}\Sigma^{\mu\nu}\Sigma_{\mu\nu} .
\end{equation}
In any astronomical system
the symmetries which hold approximatly
on a macroscopic scale are broken on a
microscopic scale. Even though the universe is isotropic on large scales the
velocities of the individual galaxies do not in general
conform with the rest frame of the cosmic microwave background.
Phenomenologically this situation can be described by
regarding
each galaxy as a particle in a cosmic fluid.
In this model the pressure is a measure of the microscopic deviation from
the perfect symmetry \cite{Soleng0}.
For a general perfect fluid the four-velocity is an {\em averaged\/}
four-velocity.
In the same way as pressure is introduced for a conventional
fluid,
one can add a stress tensor proportional to the projection tensor that
projects any direction into the space perpendicular to the
surface spanned by the {\em averaged\/} bivector $\Sigma^{\mu\nu}$.
Let this projection tensor be given by
$-\Sigma^{\mu\lambda}\Sigma_{\lambda\nu}/\gamma + \delta^{\mu}_{\;\;\nu}$, and
let
$q$ represent the ``pressure'' of the string fluid.
Then, the energy-momentum
tensor for a perfect fluid of strings is \cite{Letelier2}
\begin{equation}
T_{f}^{\mu\nu}=\left(q+\sqrt{-\gamma}\rho\right)\Sigma^{\mu\lambda}
\Sigma_{\lambda}^{\;\;\nu}/(-\gamma)+q g^{\mu\nu} .
\label{energy}
\end{equation}

\section{Exact solution}
\addtocounter{equation}{-\value{equation}}
Consider a static, spherically  symmetric space--time.
We may write the metric as
\begin{equation}
ds^2=-e^{2\mu}dt^2+e^{2\lambda}dr^2+r^2d\theta^2
+r^2\sin^2{\theta }\, d\phi^2  \label{metrikk}
\end{equation}
where $\mu$ and $\lambda$ depend on the radial coordinate $r$, only.
The symmetries of this metric
restrict $\Sigma_{\mu\nu}$ to have only two nonvanishing
components, $\Sigma_{tr}$ and $\Sigma_{\theta\phi}$.
Physically this means that the averaged world sheets are either in the
$(r,t)$ or $(\theta, \phi)$-planes.
The condition
$\gamma <0$ (the world sheets are not space-like)\
and the definition (\ref{bivector}), tell us that only the
$\Sigma_{tr}$ component survive \cite{Letelier}.
This means that the averaged world sheets are in the $(t,r)$-plane, but
only in the
exceptional
case of a string cloud does the
individual strings
orient themselves
in a perfect radial fashion.
Thus, the energy-momentum tensor (\ref{energy})
of a perfect string fluid, reduces
to
\begin{equation}
T^{t}_{\;t}=T^{r}_{\;r}\;{\mbox{ and }}\;
T^{\Omega}_{\;\;\Omega}=q
\end{equation}
where $\Omega$ stands for both $\theta$ and $\phi$.

The Ricci tensor for the line-element (\ref{metrikk})\
takes the form
\begin{eqnarray}
R^{t}_{\;\;t}&=&
\left(    -\mu''+\lambda'\mu'-\mu'^2-2\frac{\mu'}{r}
\right) e^{-2\lambda} \nonumber\\
{\mbox{ }} & & {\mbox{ }}\nonumber\\
R^{r}_{\;\;r}&=&
\left(    -\mu''+\lambda'\mu'-\mu'^2+2\frac{\lambda'}{r}
\right) e^{-2\lambda}\\
{\mbox{ }} & & {\mbox{ }}\nonumber\\
R^{\Omega}_{\;\;\Omega}&=&
\left(-\frac{\mu'}{r}+\frac{\lambda'}{r}-\frac{1}{r^2}\right)e^{-2\lambda}
+\frac{1}{r^2}. \nonumber
\end{eqnarray}

The gravitational field is uniquely determined once
an ``equation of state'' specifies the relation between
$T^{t}_{\;\;t}$ and $T^{\Omega}_{\;\;\Omega}$.
Here we shall assume that $T^{t}_{\;\;t}$ is proportional to
$T^{\Omega}_{\;\;\Omega}$.
Thus we get an
energy-momentum tensor of the form\footnote{Energy-momentum tensors of this
algebraic form (also with a variable $\alpha$)\
were first considered by Petrov \cite{Petrov},
and later discussed by Gliner \cite{Gliner}.
Dymnikova \cite{Dymnikova} and Soleng \cite{Soleng}
have recently interpreted it as the energy-momentum tensor associated with
anisotropic vacuum polarization in a spherically symmetric space--time.}
\begin{equation}
T^{t}_{\;\;t}=T^{r}_{\;\;r}=-\alpha T^{\Omega}_{\;\;\Omega},
\label{Ansatz}
\end{equation}
where $\alpha$ is a dimension\-less constant.
Using geometrized units, i.e.\ $G=c=1$, and the above
energy-momentum tensor, Einstein's field
equations
\begin{equation}
G^{\mu}_{\;\;\nu}\equiv R^{\mu}_{\;\;\nu}-\frac{1}{2}g^{\mu}_{\;\;\nu}R
=8\pi  T^{\mu}_{\;\;\nu},
\end{equation}
imply
\begin{equation}
G^{t}_{\;\;t}=G^{r}_{\;\;r}\;\;\;{\mbox{ and }}\;\;\;
G^{t}_{\;\;t}=-\alpha G^{\Omega}_{\;\;\Omega} .
\end{equation}
{}From the first equation we find $\lambda=-\mu$.
Then the second equation becomes
\begin{equation}
\frac{1}{r^2}\left[( r e^{2\mu})'-1\right]=
-\frac{\alpha}{2r}\left[( r e^{2\mu})'-1\right]' .
\label{secondeq}
\end{equation}
The case $\alpha=0$ corresponds to the Schwarzschild solution.
For $\alpha\neq 0$, integration yields
\begin{equation}
( r e^{2\mu})'-1=-\varepsilon\left(\frac{\ell}{r}\right)^{2/\alpha}
\label{secondint}
\end{equation}
where $\ell$ is a positive integration constant of dimension
length and $\varepsilon=\pm 1$ is a sign factor which determines the sign of
the
energy-density of the string fluid ($\varepsilon=1$ corresponds to
a positive energy density). Integrating once more, we find
\begin{equation}
e^{2\mu}=1-\frac{2M}{r}-\left\{  \begin{array}{lcl}
\varepsilon\ell r^{-1}\ln{(\lambda r)} &
{\mbox{ for }} & \alpha=2\\
{\mbox{ }}&
{\mbox{ }}&
{\mbox{ }}\\
\varepsilon\alpha(\alpha-2)^{-1}\ell^{2/\alpha}r^{-2/\alpha} &
{\mbox{ for }} & \alpha\neq 2.
\end{array}
\right.  \label{solution}
\end{equation}
This is the general solution for static and
spherically symmetric  space--times with a radially
boost invariant energy-momentum tensor with a constant equation of state.
The following special cases are singled out: $\alpha=-1$ corresponds to the
Schwarzschild--de~Sitter solution, $\alpha=0$
gives the
Schwarzschild solution, and  $\alpha=1$ represents the
Reissner-Nordstr{\"o}m solution.
For the case $\alpha = \infty$, the correction
term in the metric coefficient of
Eq.~(\ref{solution})\ is a constant, and thus
in the zero temperature limit the strings do not produce any
gravitational forces. This result agrees with Letelier's \cite{Letelier}
solution for a static
cloud of strings with spherical symmetry.
In this case the strings are exactly straight and radially directed.
A straight ideal string has vanishing gravitational mass, because
the gravitational effect of tension exactly cancels the
effect of its mass \cite{Gott2,Hiscock}. The only remaining gravitational
effect is an angle deficit. The same happens in the spherically symmetric
solution, but here there is a solid angle deficit.

In the generic case,  $\alpha\in\langle-\infty,\infty\rangle\setminus\{0,2\}$,
the classical gravitational
acceleration is
\begin{equation}
g=\frac{M}{r^2}+\varepsilon\frac{\ell^{-1}}{(\alpha-2)}
\left(\frac{\ell}{r}\right)^{1+2/\alpha} .
\label{graviacc}
\end{equation}

The energy-density, $\rho_{s}$, corresponding to
these solutions are found from
Eqs.\ (\ref{secondeq}) and (\ref{secondint}). Hence, using that
$8\pi\rho_{s}=-8\pi T^{t}_{\;\;t}=-G^{t}_{\;\;t}$, one finds
\begin{equation}
8\pi \rho_{s}= \frac{\varepsilon}{r^2}\left(\frac{\ell}{r}\right)^{2/\alpha}.
\end{equation}
The divergence of $T^{\mu}_{\;\;\nu}$ as $r\rightarrow 0$
signals that the model breaks down at small distances.
The singularity at the origin
is not unique for the string fluid model: also
the Schwarzschild solution of
a point mass in vacuum is a singular space--time, and the
Reissner-Nordstr{\"o}m solution of a charged
point mass has a diverging energy-momentum tensor at the origin.
The singularities
at the origin
do not invalidate these solutions as
models of the gravitational field
outside masses with charges or masses
surrounded by a
string fluid.

\section{Dark matter or modified gravity}
\addtocounter{equation}{-\value{equation}}

Einstein's theory has had remarkable success in explaining
observed and inferred gravitational phenomena. There seems to
be only one serious problem---the missing mass problem.
On large scales, the scales of galaxies and beyond, the
Einstein-Newton
dynamics
seems to imply that there is
much more mass than we can observe directly.
{}From observations of the rotational velocities of the
gaseous component of galaxies, it is found that
the velocity approaches a constant at large distances
\cite{SancisivanAlbada},
and
from the relation
$v^2/r=g$, one finds that
the gravitational acceleration decreases as $1/r$ here.
According to Newtonian gravity,
this corresponds to an effective mean mass density $\rho (r)\sim 1/r^2$ and
a total mass increasing linearly with distance.
If luminuous matter were a good tracer
of mass, and if Newton's law were valid at these scales, one should
find $g\sim 1/r^2$ and a mean mass density of $\rho(r)\sim 1/r^3$
corresponding to a constant total mass at large distances.

There are three explanations to this discrepancy: ($a$) Newtonian dynamics
is wrong at these scales, ($b$) there is a lot of unseen ``dark matter"
in the galaxies, or ($c$)
non-gravitational forces play an important r{\^o}le in the motion of
the gaseous component of galaxies, e.g.\ magnetic fields
\cite{Battaneretal}.
Many authors have
advocated the first explanation, and modifications of Newton's
gravitational dynamics have been proposed
\cite{Finzi,Milgrom83a,BekensteinMilgrom84,KuhnKruglyak,Sanders,Liboff}.
Milgrom's theory (for reviews, see Refs.\
\cite{Milgrom87,MilgromBekenstein87,Milgrom89})\
has worked impressingly well
both for galaxies \cite{Milgrom83b,Milgrom84,Milgrom86}
and galaxy systems \cite{Milgrom83c},
and Begeman et al.\ \cite{Begemanetal} have {\em claimed\/} that it gives
the best phenomenological description of
the systematics of the mass discrepancy in galaxies.
Others come to a different conclusion, cf.\ Ref.\ \cite{Lake} (see also
Milgrom's
rebuttal \cite{Milgrom91})\ and
Ref.\ \cite{GerhardSpergel}. Another argument
in favor of an effective $1/r$ correction to the force law at large
distances, is that such a term could stabilize a cold
stellar disk in a numerical galaxy model \cite{Tohline}.
But this  would happen also if the
$1/r$ effective force law was due to dark
matter.

In spite of the phenomenological success of non-Newtonian dynamics,
the scientific community has been reluctant to
abolish Newton's theory of gravitation, partly because
Newton's theory is far more aesthetically attractive
than any of the modified theories, and since the
interactions appear to
be more fundamental than matter, one would
rather introduce new matter than new forces, but above all
it is objected that none of the modified theories have a viable
relativistic counterpart \cite{Milgrom89,Lindley}.
The attempts
in this direction, the aquadratic Lagrangian theory \cite{BekensteinMilgrom84},
and phase coupling gravitation
\cite{Bekenstein88a,Bekenstein88b}
have lead to tachyonic propagation \cite{Bekenstein90}
or problems with
gravitational lensing measurements \cite{Bekenstein92},
respectively, and
it seems that if a viable
 scalar-tensor theory that mimics Milgrom's exists, it
will be very complicated and contain many new
fundamental constants.
Therefore, the most widely accepted explanation is that
the universe is filled with huge amounts of dark matter.

{}From Eq.\ (\ref{graviacc}) one gets
an effective  attractive
 $1/r$ correction to Newton's force if
$|\alpha|\gg 1$.
It is therefore tempting to propose a string fluid
as a specific form of dark
matter.
In the case of spherical symmetry, this energy-momentum tensor
re\-pro\-duces a force law similar to Milgrom's
\cite{Milgrom83a}
within the
framework of General
Relativity \cite{Another}.

\section{String fluid background and dark matter}
\addtocounter{equation}{-\value{equation}}

\subsection{String fluid cosmology}

It has earlier been proposed that dark matter on cosmic scales may be
composed of strings \cite{Vilenkin84}.  A good review of the physics
of cosmic strings can be found in Ref.\ \cite{Vilenkin85}.
The astronomical constraints on a string-dominated universe
have been analyzed
by
Gott and Rees \cite{GottRees}. They concluded that if
 the strings have a mass
per unit length below
$G\mu\approx
10^{-14}$--$10^{-15}$,
the strings
could contribute with more than five times the critical
density and still be in agreement with observational data.
If the
mass per length
is very small,
such as
$G\mu\approx 10^{-30}$,
then the strings would be
difficult to detect
even when passing through the
observer
\cite{Vilenkin84}.
A string-dominated universe
requires that the intercommuting probability of the strings
is very small, i.e., the
strings pass freely through one another.
In contrast,
simulations of string models have lead to a {\em high\/} intercommuting
probability.
On account of this,
string-dominated universe models have lost favour
among theorists. Needless to say,
the evolution of a string network
is model dependent,
and just as one has very weakly interacting
particles, it is {\em a priori\/} possible to have very weakly interacting
strings. Hence, there is no fundamental principle
which forbids
a string-dominated universe. Also, astronomical data are compatible
with this hypothesis. Admittedly, it could be discussed
how {\em natural\/} this hypothesis is, but for the moment we
adopt it as a working hypothesis in order to see where it leads.

An isotropic stringy background (a string cloud)\
has an energy-density of the form \cite{Vilenkin85}
\begin{equation}
8\pi \rho_{s}=\frac{3w}{R^2}
\end{equation}
where $R$ is the cosmic scale factor and $w$ is a constant.
Accordingly, the first Friedmann equation takes the form
\begin{equation}
H^2= \frac{8\pi}{3}\rho_{m}+\frac{w-k}{R^{2}}
\end{equation}
where $\rho_{m}$ represents the non-stringy energy density.
This means that if one neglects the stringy
background, the effective curvature is
$k_{{\mbox{{\footnotesize{eff}}}}}=k-w$ rather than $k$.
In this way a closed universe with a geometric $k\approx 0$
as predicted by inflation
could be in agreement with the observed
$k_{{\mbox{{\footnotesize{eff}}}}}<0$, if the universe
is filled with a string cloud \cite{Vilenkin85}.

\subsection{String fluid in galaxies}

The hypothesis that
the universe is
filled with a string fluid where each string has a mass density
below what is directly observable,
can only be tested by the effect of the
string fluid
on the local gravitational fields of galaxies.
On the other hand the gravitational fields of galaxies
are expected to affect the motion of strings. According to the
equivalence principle
all masses fall in the same fashion. In general,
due to tidal forces, elongated
objects will tend to orient themselves
along the gravitational field lines.
This could provide a mechanism whereby
strings are pulled
out
of the cosmic ensemble by every galaxy and redirected so as
to appear approximately radially oriented near each galaxy.
If the string tension is not too large,
it is
possible that a significant number of strings are
trapped by the galaxies. Such stringy halos could produce
a measureable change in
the long range gravitational
fields of the galaxies.

The cosmic string fluid is assumed to be cold.
Strings which are trapped by galaxies would be heated due to
release of potential gravitational
energy. Let a string segment of length
$R_{s}$ be trapped in this way. The
potential
gravitational
energy inside a region of radius
$R_{s}$ is
\begin{equation}
E_{p}=-\frac{M m_{s}}{R_{s}^2}
\end{equation}
where
$M$ is the mass of the galaxy and where
$m_{s}=\mu R_{s}$ is the mass of the trapped string
segment.
If the string was trapped in the early universe and if the
released energy flows along the string with the speed of
light, it would now be distributed over at most one Hubble
length, $H_{0}^{-1}$. Had the string been
trapped
later, this length
would have been
shorter. Let us, however, consider the Hubble length
which is relevant if most of the strings were
trapped
before or at the time of galaxy formation.
The fraction $R_{s} H_{0}$ of $-E_{p}$
remains in the trapped part of the string.
Thus, this part of the string
could gain an
  energy per mass
 of
\begin{equation}
\frac{\Delta E}{m_{s}}
\approx
-\frac{E_{p} R_{s} H_{0}}{m_{s}}
%{\;\;\mbox{\raisebox{0.5ex}{$<$}\hspace{-1.8ex}\raisebox{-0.5ex}{$\sim$}}\;\;}
=
MH_{0}
\label{gravenergy}
\end{equation}
by falling into a galaxy of mass $M$.
This energy is available to produce transverse motion of the string.
For the string
fluid this leads to a transverse pressure given by
\begin{equation}
(pV)^2=(m_{s}+\Delta E)^2-m_{s}^2 .
\label{press}
\end{equation}
The transverse pressure will
contribute with a pressure per energy density
$pV/m_{s}
%{\;\;\mbox{\raisebox{0.5ex}{$<$}\hspace{-1.8ex}\raisebox{-0.5ex}{$\sim$}}\;\;}
\approx
(MH_{0})^{1/2}$.
This determines the dimensionless constant in Eq.\
(\ref{Ansatz})\ as follows
\begin{equation}
\alpha
%{\;\;\mbox{\raisebox{0.5ex}{$<$}\hspace{-1.8ex}\raisebox{-0.5ex}{$\sim$}}\;\;}
\approx
\left(MH_{0}\right)^{-1/2}
\label{alphavalue}
\end{equation}
where we have used that the string fluid energy-density and
transverse
pressure are positive to conclude that $\varepsilon=1$.
The string fluid
contributes with a positive effective gravitational
mass. This agrees with the General Relativity
lore that pressure contributes to
gravitational attraction.
The value of $\alpha$ in Eq.~(\ref{alphavalue})\ is large.
A large $\alpha$ corresponds to
$|T^{t}_{\;\;t}|=|T^{r}_{\;\;r}|\gg |T^{\Omega}_{\;\;\Omega}|$.
Physically, the solution (\ref{solution})\ with
a large $\alpha$ represents
a fluid of
{\em nearly radially directed strings at low but nonzero temperature.\/}
Recall that the $\alpha\gg 1$ solution implies
an $1/r$ correction to Newton's
gravitational force law. This  is what is
needed to explain the flat rotation curves of galaxies.
The term $\ell^{2/\alpha}$ in Eq.~(\ref{graviacc})
is an {\em a priori\/} arbitrary normalization
constant,
that determines the
density of strings
or, equivalently, normalizes the energy density of
the string fluid. From a mathematical perspective
$\ell^{2/\alpha}$ is a free
parameter,
but from a physical point of view, certain
values are more reasonable than others: $\ell$ is a constant of
dimension length and in a fundamental underlying theory it has to be
determined by physical constants. The smallest and largest units of length
we can think of are the Planck length and the Hubble length, and
in my opinion
it is reasonable to assume
that $\ell$ is within this range.
The predicted  value of $|\alpha|$ is so large that
$(\ell/r)^{2/\alpha}\approx 1$
for both the extreme possibilities for $\ell$.
Hence,  with this information, Eq.\
(\ref{graviacc}) implies that
Newton's force law is changed to
\begin{equation}
g=\frac{M}{r^2}+k_{0}\frac{(MH_{0})^{1/2}}{r}  \label{forcelaw}
\end{equation}
where the constant $k_{0}\approx 1$.
The presence of the
Hubble parameter in the local force law signals that the
translational invariance of the background string fluid
is broken not only in spatial directions
but also in the time-direction.

Note that the force law of Eq.\ (\ref{forcelaw})
agrees with the Tully-Fisher law \cite{Tully} which
relates the rotational velocity, $v$, to the luminosity, $L$,
by $v\propto L^{1/4}$ if the luminosity is proportional to the
Newtonian mass. This is a reasonable assumption
if the ratio of dark and luminous matter densities is a constant.
Also, the mysterious coincidence that Milgrom's critical
acceleration is equal to the
Hubble parameter \cite{Milgrom83a,Milgrom89},
is explained as a result of having
dark matter of cosmic extension.

\section{Discussion}
\addtocounter{equation}{-\value{equation}}

An exact
static, spherically symmetric
solution
of the field equations of
General Relativity
coupled to a string fluid with a constant equation of state
has been found.  For the same space--time symmetries,
this is also the most general solution for
any radially boost invariant
energy-momentum tensor with a constant equation of state.
The solution is singular at the origin. This
singularity can be attributed to the
high degree of symmetry of the model, and does not
affect the behaviour at large $r$.
It has been observed that the solution
reproduces a force law similar to Milgrom's \cite{Milgrom83a}.
This motivates a string model for dark matter.
According to this
model, the new solution
can explain the Machian character of
Milgrom's acceleration, $a_{0}\approx H_{0}$.
The dark matter model presented here
is a phenomenological model,
but it can no longer be objected that
the $1/r$ modification of Newton's force law is ``an orphan in the classical
world'' \cite{Lindley}.
Instead
it follows from General Relativity with
an energy-momentum tensor corresponding to
a string fluid
provided this string fluid satisfies the equation of state
$T^{t}_{\;\;t}=-\alpha T^{\Omega}_{\;\;\Omega}$ where
$\alpha\approx (MH_{0})^{-1/2}$.
Reasonable assumptions concerning the infall of strings
into galaxies have lead to this value for $\alpha$.

Up to now the missing mass problem has been resolved by assuming that
dark matter is present in whatever quantities and distributions that are
needed to explain away all mass descrepancies. The main problem with this
approach is that the dark matter hypothesis in this form
is too flexible to give
any unavoidable predictions \cite{Milgrom89}, and it is in principle not
testable before
one specifies the nature of the dark matter.
In contrast,
the approach of Milgrom is testable, and it
gives specific predictions which are in good agreement with
observations. The main problem has been that the modified dynamics
has had no viable relativistic counterpart.

Despite the success of Milgrom's $1/r$ force law and the fact that the present
model reproduces it, there are many reasons
why it is premature to identify a string fluid as
{\em the solution\/} to the missing mass problem. First, there is no
field theoretic realization of this particular model.
Second,
the value $\alpha\approx (M H_{0})^{-1/2}$
has been found only with the additional assumption that the
potential gravitational energy
of the strings has been spread over a length of $H_{0}^{-1}$.
On the other hand, it is natural to suspect
that an isotropic background of strings
which only results in a redefinition of the effective spatial
curvature on cosmic scales, only contributes with a similar
redefinition of the effective asymptotic space in the geometry of
isolated masses. Then,
only the gravitationally trapped strings, i.e.,
those described by the energy-momentum tensor of the solution,
are relevant
for local gravitational field, and the solution is in fact unique
up to a specification of $\alpha$.
For $\alpha^{-1}$ which is zero in the string cloud of the cosmic
background, one would expect only a small perturbation near galaxies.
The perturbation away from $\alpha^{-1}=0$
would be expected to depend on the two characteristic length
scales of the problem, namely the Hubble parameter
and  the mass of the galaxy.
This is just the result obtained by an energy argument.
To clarify this issue one needs to solve the embedding problem
where the local solution is embedded in a string-dominated
universe.
Third,
the proposal has only been studied in a static, spherically symmetric
model.
In contrast, the real universe is non-static and it contains
many galaxies, so both of the symmetry assumptions are broken.
The consequences of
deviations from spherical
symmetry are not clear, and it is also a question how
strings passing through more than one galaxy will affect
the model.
Finally, one expects realistic string models to predict
that strings will intercommute and produce closed loops
which collapse and leave the universe with very few strings.

In spite of these problems,
from a general relativistic perspective, it is very interesting
that such a simplistic model can reproduce the non-Newtonian force law.
The generally accepted solution---dark matter---need therefore not be
a very complicated system of epicycles as have been argued by
the proponents of non-Einsteinian gravitation, e.g.\ Sanders \cite{Sanders}.

\subsection*{Acknowledgements}
It is a pleasure  to thank
Paul J. Steinhardt
for stimulating my interest in the dark matter problem and
for
valuable comments.
I am also grateful to
Jacob Bekenstein
and
Mordehai Milgrom
for correspondence and critical remarks, and I thank
{\O}yvind Gr{\o}n for carefully
reading through the manuscript.
Thanks are also due to the referees who have contributed with
constructive criticism and a number of references to the
literature.
This research was initiated at the University of Pennsylvania where
I spent
my sabbatical year 1992--93
on leave from from the University of Oslo. I
acknowledge support from
the Thomas Fearnley Foundation,
from Lise and Arnfinn Heje's Foundation, Ref.\ No.\ 0F0377/1993,
and from the U.S.\ Department of Energy under Contract
No.\ DOE-EY-76-C-02-3071.

	%              REFERENCES

\end{document}